\documentclass{ws-mpla}
\usepackage{graphicx}
\usepackage[usenames,dvipsnames]{color}
\usepackage{amsmath,amssymb,amsfonts,bm}
\usepackage{hyperref}

\begin{document}

\markboth{Tamal K. Mukherjee,Soma Sanyal}
{Particle temperature and the Chiral Vortical Effect in the early universe.}

\title{Particle temperature and the Chiral Vortical Effect in the early universe.}

\author{\footnotesize Tamal K. Mukherjee
\footnote{tamalk.mukherjee@visva-bharati.ac.in}}
\address{Department of Physics, Visva Bharati, Santiniketan 731235, India}
\author{\footnotesize Soma Sanyal
\footnote{sossp@uohyd.ernet.in}}
\address{School of Physics, University of Hyderabad, Telangana, 500046, 
India}

\maketitle

\begin{abstract}

We study the effect of hotter or colder particles on the evolution of the chiral magnetic field in the early universe. We are interested in the temperature dependent 
term in the chiral vortical effect. There are no changes in the magnetic energy spectrum at large lengthscales but in the Kolmogorov regime we do find a 
difference. Our numerical results show that the Gaussian peak in the magnetic spectrum becomes negatively skewed. The negatively skewed peak can be 
fitted with a beta distribution.  Analytically one can relate the non-Gaussianity of the distribution to the temperature dependent vorticity term. 
The vorticity term is therefore responsible for the beta distribution in the magnetic spectrum. Since the beta distribution has already been used to model 
turbulent dispersion in fluids, hence it seems that the presence of hotter or colder particles may lead to turbulence in the magnetized plasma.

\end{abstract}

\ccode{PACS Nos: 98.80.-k,98.80.Cq,95.35.+d}

\section{INTRODUCTION}
\label{sec:intro} 

There are magnetic fields everywhere in our universe. Hence the generation and evolution of magnetic fields in our universe have been extensively studied \cite{Grasso:2000wj}. 
Observational data in the intergalactic space indicate that magnetic fields will have a primordial origin. These then evolve with the expansion 
of the universe to give the fields that are observed today. It is now well accepted that seed fields were generated in the early universe, by various 
mechanisms \cite{Widrow:2002ud,Widrow:2011hs}. They may be the consequences of charge separation by turbulence during cosmological phase transitions or from 
various inflationary scenarios. 
These seed fields are small but they are amplified by various dynamo mechanisms as the universe evolves. Unfortunately, existing data cannot put 
direct constraints on the properties and magnitude of the seed magnetic fields. This means that the generation and evolution of the initial field still remains 
an open problem in cosmology \cite{Durrer} .

Recently, studies have shown that chiral effects may play a significant role in the evolution of the magnetic fields in the early universe. The chiral 
asymmetry of the cosmic plasma  has been discussed in relation to baryogenesis and leptogenesis scenarios in the early universe. The presence of a 
chiral asymmetry can also lead to the generation of a seed magnetic field \cite{long,sydorenko,anand}. There are various effects that chirality has on the generation 
and evolution of a helical magnetic field in the early universe. Amongst them, the chiral magnetic effect and the chiral vortical effect have been studied 
\cite{tashiro,ruchaivsky} extensively. Let $\mu_L$ and $\mu_R $ be the chemical potential of the left handed and the right handed particles in the early universe.
In the presence of a magnetic field, a non zero current is generated, which is proportional to the product of the magnetic field and $\Delta \mu = (\mu_L - \mu_R)$. This is 
the chiral magnetic current \cite{kharzeev,fukushima}. Similarly, in the presence of vorticity and chiral asymmetry, a current develops which is proportional to 
the vorticity and $\Delta\mu^2$. This is the chiral vortical effect \cite{tashiro,son}. Both these effects have been shown to 
affect the evolution of a helical magnetic field.

Now, one of the problems in the models of generation of primordial fields is that they are not strong enough to survive the expansion of the universe.
The generation of helical magnetic fields and the presence of chiral asymmetry during the evolution of the fields slows down the decay of the magnetic fields. 
There have been many detailed studies of the generation and the evolution of helical magnetic fields in the universe \cite{tashiro,tina}. These studies show that 
the magnetic helicity has an attractor behaviour with respect to the chemical potential difference. No matter what is the initial value of $\Delta\mu$, 
they all evolve to decay along a common trajectory. This has a slope of $T^{1/2}$ \cite{joyce}. Eventually, the magnetic helicity reaches an asymptotic value and becomes 
maximally helical. Chiral effects also produce a Gaussian peak in the magnetic spectrum due to the dominance of the chiral magnetic effect over the chiral 
vortical effect \cite{tashiro}.

All these evolutions have been discussed for the cases when all the particles are in thermal equilibrium at the same temperature. Now, there also exists the 
possibility of generating thermal particles at a different temperature than the background plasma. There are various mechanisms which generate these hotter (or colder)
particles \cite{falkowski}. They can be generated at the end of inflation during the preheating stage due to parametric resonance of the oscillating inflaton field. They can be generated by the 
decay of unstable topological defects. Some of the these particles may be chirally asymmetric. Examples of such particles are non-thermal right - handed 
neutrinos which may or may not equilibrate during the history of the universe \cite{chen}. Vilenkin \cite{vilenkin} had obtained the net current of neutrinos 
from rotating black holes and found that the net current depends on both the asymmetry as well as the temperature. For electrons, the net current exists in a 
nonequilibrium situation.

Generally it is assumed that both the right handed and the left handed particles are in thermal equilibrium at a temperature $T$. But if there is a 
particle at a different temperature, then the chiral vortical current will have an additional term \cite{vilenkin}. We will show that such scenarios may arise 
in the early universe. However, instead of focussing on one such particular scenario,
we will try to understand the effect of the additional term on the evolution of the magnetic fields in general. We find that the presence of a particle at a 
different temperature, will result in the generation of strong magnetic fields in the early universe provided the particle has been generated at a higher 
temperature. Even though, strong magnetic fields are generated, the chiral magnetic effect does not dominate over the chiral vortical effect. The hotter/colder 
particles will keep injecting energy in the system. We find that instead of the Gaussian peak in the magnetic spectrum, a negatively skewed peak is observed. 
This peak can be fitted by a beta distribution. Since the family of beta distributions has been previously linked to turbulence, it looks like the hotter/colder particles 
may lead to turbulence in the plasma. Turbulence in the early universe and its effect on the chiral magnetic effect has been studied previously 
Dvornikov et. al \cite{dvornikov}.

In section 2, we discuss the generation of particles at a different temperature. In section 3, we discuss the chiral vortical effect and 
the current due to the chiral vortical effect.  We then discuss the evolution of a helical field in the early 
universe in the presence of these particles in section 4. In section 5, we present the numerical results of the magnetic field evolution equations
and show that the peak can be fitted with a beta distribution.  Finally in section 6, we summarize and discuss the results obtained.

\section{Generation of particles at a different temperature }
 \label{sec:hotparticle}
Usually particles in the early universe are in thermal equilibrium with the background plasma. However, due to various mechanisms, particles may be generated which 
may not be in thermal equilibrium with the background plasma. Most of these are dark matter candidates. Some examples are non-thermal axions produced by vacuum 
alignment, winos produced from moduli decay etc \cite{acharya}. One of the most popular methods of non thermal generation of particles is during reheating 
after inflation. The rapid oscillation of the inflaton field towards the end of inflation generate non - thermal gravitinos \cite{giudice}. 
Non - thermal right handed neutrinos are also produced by these kind of oscillations. These may not thermalize during the entire history of the universe. 
Non thermal particles such as neutralinos may also be generated during decay of topological defects \cite{jean}. At this point, we wish to clarify that we 
are interested in only one aspect of these non-thermal particles. Their temperature should be different from the background temperature. Since we 
are not interested in any other aspects of these particles we refer to them as hotter or colder particles. However, in this section, since we are discussing
the work of other groups we will refer to these particles as non-thermal. In the rest of the sections, we will refer to them as hotter/colder particles.

 We point out a recent model by Mu-Chun Chen et al \cite{chen}. They show that it is  possible to have a non-thermal neutrino background 
with comparable background density as the thermal neutrino background. The non-thermal neutrino background is generated in a model of chaotic inflation during 
inflationary preheating. Generally, in the case of preheating via parametric resonance, the momenta of the fermions are distributed stochastically over a sphere of 
radius $\epsilon_F$ (the fermi energy). This itself depends upon a parameter $q$ which is the resonance factor given by $q = \frac{\lambda^2 \phi_{0}^2}{m_{\phi}^2} $. 
Since $q$ depends on the couplings in the underlying theory, it is possible to obtain a non thermal spectrum of neutrinos at a temperature higher than the reheating 
temperature. It requires the presence of a term, 
\begin{equation}
 \mathcal{L} \supset \lambda \phi \bar{\nu_R^C} \nu_R + h.c.
\end{equation}
As long as this term is allowed and the inflaton is appropriately charged to prohibit the Majorana mass term, a nonthermal neutrino background can be generated.
These are right handed neutrinos. They are relativistic and it has been shown that they do not thermalize either above or below 
the electroweak scale. The energy of the particles are however red-shifted and they do get diluted  by the expansion of the universe. At very high temperatures above the 
electroweak scale it is possible that such particles exist. Their temperature will not be equal to the thermal background temperature. In fact, they may have a temperature 
higher than the background temperature. There are other models where charged leptons, neutralinos etc. can have a nonthermal distribution in the early universe 
\cite{asaka,covi,dermer}. There are extensions of the standard model where chiral fermions are predicted \cite{patra,basso} as dark matter candidates. It is possible 
that these are nonthermal fermionic dark matter with charges that are different from the standard model fermions \cite{biswas}. The important thing that we 
want to establish here is that it is possible that though the early universe plasma is in thermal equilibrium, some species of particles in the plasma may be at a higher 
temperature. The number density of the particles will then vary and if the particles are charged, it will lead to an increase in the net current in the plasma due to the chiral 
vortical effect. 

The early universe plasma has a very high Reynolds number and therefore magnetic fields can be generated due to the vorticity in the plasma. We are however
not interested in the generation of magnetic fields. Magnetic field generation due to the abelian anomaly and chiral vorticity has been discussed 
before in different contexts\cite{joyce,bhatt}. Any of these mechanisms which generate the magnetic field will lead to some vorticity in the plasma. 
Usually the vorticity gradually dies down as the strength of the magnetic field increases. The dissipative forces and the chiral magnetic effect takes over 
and vorticity can gradually be neglected. However, in the presence of particles at a different temperature, we show that even though the magnetic field increases, the CVE
cannot be neglected. Thus vorticity in the plasma remains and subsequently the plasma may become turbulent.

\section{The Chiral Vorticity Effect}
\label{sec:cve}
Generally in the presence of charged fermions one can obtain an electrical current if the number density of the positively charged particles are not equal to
the negatively charged particles. Vilenkin \cite{vilenkin} showed that a vector current can
be generated in the presence of a chiral imbalance. A chiral imbalance implies 
that the number density of the right handed particles are not equal to the 
left handed particles. In the context of a rotating black hole, he showed that the presence of vorticity and chiral imbalance gives rise to a current given by 
\begin{equation}
 \label{cvecurrent1}
 {\bf{J}}_{\chi \omega} = \frac{e}{4 \pi^2} \Delta \mu^2 {\bm{\omega}} +  \frac{e}{12} T^2 {\bm{\omega}}
\end{equation}
Here, $e$ is the electric charge and $\omega$ is the vorticity in the plasma. 
Though the net neutrino current density was calculated in the paper, it was argued that a similar current will arise in the case of other particles which are 
found in both negative and positive helicity states. 
In equation (2) the temperature $T$ is the temperature of the particles being emitted from the black hole. It is not the background temperature. Subsequently, 
for a general case of particles in an equilibrated plasma, since all the particles are at the same temperature, the term was usually dropped from the chiral vortical current.
Thus the chiral vortical current is modified when the left handed and right handed particles and anti particles are 
in thermal equilibrium at the same temperature. It is then given by, 
\begin{equation}
 \label{cvecurrent2}
 {\bf{J}}_{\chi \omega} = \frac{e}{4 \pi^2} \Delta \mu^2 {\bm{\omega}}
\end{equation}

Later it was shown that apart from the vorticity effect arising from the rotation of the plasma, 
there are several other
effects which arise due to the chiral imbalance in the plasma. Especially in the presence of
a magnetic field, there occurs the Chiral Magnetic Effect (CME). The CME arises not 
only from the interplay of the chiral imbalance and the magnetic field, but it is related to the chiral anomaly present in the underlying theory. Both the CVE and 
the CME has been discussed in detail with applications 
both in the early universe as well as in heavy ion collisions \cite{son,kharzeev2,boyarsky}. We refer the readers to some of these reviews for a more detailed discussion on these two effects. 
The evolution of magnetic fields under these effects has also been studied by Tashiro et.al.,\cite{tashiro} in the context of the early universe plasma. They had found that any 
magnetic field evolves to a maximally helical configuration due to these effects. They also found a strong narrow Gaussian peak in the magnetic spectrum. This peak shifts to longer 
lengthscales and became sharper with the expansion of the universe.

\section{Evolution of helical magnetic fields in the early universe}

The general evolution of a magnetic field generated in the early universe depends on its strength, spectrum and helicity content. The strength of the magnetic fields 
generally decreases with the expanding universe. The magnetohydrodynamic equations are evolved in a Robertson-Walker metric, 
\begin{equation}
 ds^2 = R^2(\eta) (-d\eta^2 + \delta_{ij} dx^{i} dx^{j})
\end{equation}
where $R(\eta)$ ($R \, = \, \frac{1}{T}$, $T$ being the cosmic temperature) is the scale factor. The equations are usually defined in conformal time $\eta $ with comoving variables. The conformal time is defined by 
$\eta = \sqrt{\frac{90}{8 \pi^3 g}} \frac{M_p}{T}$. Here $g$ ($g \sim 100$)
 represents the effective degrees of freedom at the epoch being considered (i.e,radiation epoch). All the Maxwell's equations can then be 
defined using the comoving variables. Quite conveniently, the form of the Maxwell's equations remain the same in the
comoving reference frame only the variables change to their respective comoving variables. So the magnetic field is now given by $B = R^2(\eta)B(\eta)$ and similarly 
the electric field and the other quantities are also redefined. The Maxwell's equations are given by,   

\begin{equation}
{\bf \nabla \times B} = {\bf J}, 
\label {eqB}
\end{equation}
\begin{equation}
 {\bf \nabla \times E} = \partial_\eta {\bf B}. 
 \label {eqE}
\end{equation}
The total current (${\bf J}$) here is given by the contributions from the Ohmic 
(${\bf {J}}_{ohm}$) as well as the chiral vortical (${\bf {J}}_{cve}$) 
and the chiral magnetic (${\bf {J}}_{cme}$) currents. The current arising from the chiral vortical effect has been discussed in
detail in the preceding section and for all the particles in thermal equilibrium at the same temperature it is given by \cite{tashiro} equation (3). 
When we have the presence of some particles at a different temperature $T_{NT}$, the equation gets modified to include the temperature dependent 
term and is given by, 
\begin{equation}
  {\bf{J}}_{\chi \omega} = \frac{e}{4 \pi^2} \Delta \mu^2 {\bm{\omega}} +  \frac{e}{12} T_{NT}^2 {\bm{\omega}} . 
\end{equation} 
The Ohmic component of the current has the usual form given by
\begin{equation}
	{\bf {J}}_{ohm} \, = \, \sigma \left( {\bf E} \, + \, {\bf v} \times {\bf B}  
\right),
\end{equation}
where ${\bf E}$ and $\sigma$ are the electric field and the electrical
	conductivity of the plasma. The current due to chiral magnetic effect can be written
as \cite{kharzeev}
\begin{equation}
	{\bf {J}}_{cme} \, = \, \frac{e^2}{2 \pi^2} \Delta \mu \, {\bf B},
\end{equation}
where $\Delta \mu$ is the difference in the chemical potential of the right
handed and left handed particles.
Using equations (\ref{eqB}), (\ref{eqE}) together with the expression for the
currents, the evolution  of the magnetic field can be written as
\begin{equation}
\label{mhd}
 \partial_\eta {\bf B} = {\bf \nabla} \times ({\bf v} \times {\bf B}) 
+ \gamma_D \nabla^2 {\bf B} + \gamma_{\omega} {\bf \nabla} \times {\bm\omega} 
+ \gamma_B {\bf \nabla} \times {\bf B} 
\end{equation}
where $ \gamma_D = \frac{1}{\sigma}$ and $ \gamma_B = \frac{e^2 \Delta\mu}{2 \pi^2 \sigma}$.
The vorticity term $\gamma_\omega$ is given by $\gamma_\omega = \frac{e \Delta \mu^2}{4 \pi^2 \sigma}$, when the magnetic field evolution is studied for all 
the particles in thermal equilibrium at the same temperature  \cite{tashiro}. However, in the presence of a particle with a different temperature ($T_{NT}$), 
there is an additional temperature dependent term.  It is then given by $\gamma_\omega = \frac{e \Delta \mu^2}{4 \pi^2 \sigma} + \frac{eT_{NT}^2}{12 \sigma}$.

The derivatives here are all with respect to conformal time and comoving lengths. The conductivity is the comoving electrical conductivity. A qualitative picture of the 
primordial magnetic field can be obtained by solving this equation. Due to the inherent non-linearity in the evolution equation, it is difficult to obtain an exact solution 
both numerically and analytically, hence some assumptions are made. Also it is more convenient to solve the equations and study the magnetic field evolution in terms of the
Fourier components. The standard way in which these fields are analyzed is by mode expansion.  The homogeneity and isotropy of space implies that the wavenumber ($k$) space is also homogeneous and isotropic. Hence the
spectrum of the magnetic field in wavenumber space will depend only on the wave vector $k$ and $\delta_{ij}$ and $\epsilon_{ijm}$. The spectrum of the magnetic field is then given by, 
\begin{equation}
 \langle B_i(k,\eta) B_j^{*} (k',\eta) \rangle = (2 \pi^3) \delta({\bf {k - k'}}) [(\delta_{ij} - \hat k_{i} \hat k_{j}) P_B(k) - i \epsilon_{ijm} \hat k_m P_H(k)], 
\end{equation}
where $\hat k = \frac{\bf k}{k}$ and $k = |k|$. The bracket $\langle \rangle $ denotes the ensemble average. $P_B (k)$ and $P_H (k)$ denote the symmetric and the anti-symmetric 
part of the magnetic power spectrum. A right handed orthonormal system (${\bf e^{(1)}}, {\bf e^{(2)}}, {\bf k}$) is chosen and the different fields are decomposed into divergence free and curl free fields.
The magnetic field is then given by 
\begin{equation}
 B(k, \eta) = B_{+} (k,\eta) \, {\bf e^{+}} +  B_{-} (k,\eta) \, {\bf e^{-}}
\end{equation}
where ${\bf {e^{(+)}}} \, = \, ({\bf {e^{(1)}}} + i {\bf {e^{(2)}}})/2$ and
${\bf {e^{(-)}}} \, = \, ({\bf {e^{(1)}}} - i {\bf {e^{(2)}}})/2$.
Therefore we have, 
\begin{equation}
 (2 \pi^3) \delta({\bf {k - k'}}) P_B(k) = \langle B_{+}(k) B_{+}^{*} (k') \rangle + \langle B_{-}(k) B_{-}^{*} (k') \rangle
\end{equation}
and, 
\begin{equation}
 (2 \pi^3) \delta({\bf {k - k'}}) P_H(k) = \langle B_{+}(k) B_{+}^{*} (k') \rangle - \langle B_{-}(k) B_{-}^{*} (k') \rangle
\end{equation}
One can then define the energy density as, 
\begin{equation}
 E_B = \frac{1}{2} \langle B(x) B(x) \rangle = \frac{1}{2(2\pi)^6} \int {d^3 k d^3 k' \langle B(k,\eta) B^{*}(k',\eta)\rangle e^{i{\bf x}. (\bf k - k')}} 
\end{equation}
However, for our calculation of the energy spectrum, we will be using the energy density per unit log interval which is given by, 
$\frac{d E_B}{d (log k)} = \frac{k^3 P_B}{2 \pi^2}$. 
Since the evolution of the magnetic field also depends on the velocity, we need to define the power spectrum of the velocity field too. The power 
spectrum of the velocity field is given by, 
\begin{equation}
 \langle v_i(k,\eta) v_j^{*} (k',\eta) \rangle = (2 \pi^3) \delta({\bf {k - k'}}) [(\delta_{ij} - \hat k_{i} \hat k_{j}) P_{vs}(k) - i \epsilon_{ijm} \hat k_m P_{va}(k)], 
\end{equation}
Here the velocity field is solenoidal and the fluid kinetic energy density is given by, 
\begin{equation}
 E_v = \frac{\rho}{2} \langle v(x) v(x) \rangle = \rho \int {\frac{k^2 dk}{(2 \pi)^2} |v_{+}(\eta,k)|^2 + |v_{-}(\eta,k)|^2 } 
\end{equation}
where $\rho$ is the radiation density.
The fluid helicity and the magnetic field helicity can be obtained in similar ways (for details please see \cite{tashiro}).

We then write the magnetic field evolution equation Eq.(\ref{mhd}) in terms of the mode expansion. To obtain the evolution of the energy spectrum as well as 
the helicity spectrum it is imperative to find a correlation between the magnetic field and the velocity. In this we follow the argument given by Tashiro et.al., \cite{tashiro}. 
We briefly reiterate it here for completeness sake. The mode expanded magnetic field evolution equations are, 
\begin{equation}
 \partial_\eta |B_{+}|^2 = 2(-\gamma_D k^2 +\gamma_B k)|B_{+}|^2 + 2\gamma_{\omega} k^2 \langle B_{+}^{*} v_{+} \rangle,  
\end{equation}
\begin{equation}
 \partial_\eta |B_{-}|^2 = 2(-\gamma_D k^2 -\gamma_B k)|B_{-}|^2 + 2\gamma_{\omega} k^2 \langle B_{-}^{*} v_{-} \rangle,  
\end{equation}
In the evolution equations, we see that the vorticity term is the only term that can generate a magnetic field. The chiral magnetic term can only affect the evolution 
of an existing magnetic field. Assuming that it is the vorticity which generates the magnetic field we have, 
\begin{equation}
 B_{\pm} (\eta, k) = \int_{\eta_0}^{\eta} \gamma_\omega k^2 d\eta' v_{\pm}(\eta',k).
\end{equation}
To find the correlation between this magnetic field and the velocity at any other later time, we need to find out how the velocity is correlated over different times. 
For this we need to define the unequal time correlator for the velocity field ($ \langle v_{\pm}(k,\eta') v_{\pm}^{*} (k',\eta) \rangle $).

Now the velocity fields are correlated on the eddy turnover timescale $\frac{2\pi}{k v(\eta,k)}$ for any particular $k$ mode. At any timescale longer than 
the eddy scale the fields are uncorrelated. If $v(\eta,k)$ is the fluid velocity at the lengthscale $\frac{2 \pi}{k}$, this suggests we look for a function that satisfies,  
\begin{equation}
 \langle v_{\pm}(k,\eta) v_{\pm}^{*} (k',\eta) \rangle =  (2 \pi^3) \delta({\bf {k - k'}}) \langle  v_{\pm}(k,\eta)^2 \rangle 
\end{equation}
for $|\eta - \eta'| < \frac{2 \pi}{k v(\eta,k)}$  and 
\begin{equation}
 \langle v_{\pm}(k,\eta') v_{\pm}^{*} (k',\eta) \rangle = 0 
\end{equation}
for $|\eta - \eta'| > \frac{2 \pi}{k v(\eta,k)}$. 

This then gives the magnetic field velocity correlation as, 
\begin{equation}
 \langle B_{\pm}^{*}(\eta,k) v_{\pm}(\eta,k') \rangle = \gamma_\omega k^2 f(\eta,k) |v_{\pm}|^2 (2 \pi^3) \delta^3 ({\bf {k - k'}}) 
\end{equation}
where,
\begin{equation}
 f(k,\eta) = S \frac{2 \pi}{k v} tanh \left(\frac{kv}{2 \pi S} (\eta -\eta') \right)
\end{equation}
where $S$ is the fudge factor taken to be one. This ensures that the fields are correlated over the eddy turnover scales.
The last term in the evolution equations then becomes, 
\begin{equation}
 2\gamma_{\omega} k^2 \langle B_{\pm}^{*} v_{\pm} \rangle =  2\gamma_{\omega} k^4 f(k,\eta) |v_{\pm}|^2
\end{equation}
This form of the correlator is valid as long as the chiral vortical effect is dominant. Since we are interested in studying the effects of the chiral vortical 
effect all throughout the evolution hence we take this form of the correlator for our evolution equations. 

We still need a velocity distribution for the cosmic fluid. Since we are looking at the vorticity term, as mentioned before,
we assume that the magnetic field is generated at a time $\eta_i$ due to some vortical effect. In the spectral form of the Navier-Stokes equation, energy is transferred between 
two wave numbers as long as they obey the selection rules. Kolmogorov, stated that between certain $k$ ranges, the viscous dissipation of energy can be neglected.
This range is called the inertial range. In an expanding universe, this range has to be related to the Hubble time to account for the Hubble expansion. The inertial 
wavenumber therefore depends on the velocity at the time $\eta$ and is given by $k_{\nu} (\eta) = \frac{2 \pi}{v_{\nu} (\eta) \eta}$. Within the inertial range, 
Kolmogorov gave the velocity distribution as, 
\begin{equation}
	v(k, \eta) = v_{\nu} (\eta_i) \left(\frac{\eta_i}{\eta}\right)^{11/15} \left(\frac{k}{k_{\nu}(\eta_i)}\right)^{-\frac{1}{3}}
\end{equation}
We use the Kolmogorov distribution in the inertial range. Since it is known that in the radiation era, on large lengthscales, the velocity does not change. Hence 
for $k< k_{\nu} (\eta)$, we take it to be the white noise spectrum.
\begin{equation}
	v(k, \eta) = v_{\nu} (\eta_i)  \left(\frac{k_{\nu}(\eta)}{k_{\nu}(\eta_i)}\right)^{\frac{3}{2}}
\end{equation}
Althroughout we have assumed no kinetic helicity in the velocity flow so we can take $|v_{+}|^2 = |v_{-}|^2 = \pi k^{-\frac{3}{2}}|v|^2 $. We can now write down the final evolution equations for the magnetic field as, 
\begin{equation}
\label{final1}
 \partial_\eta |B_{+}|^2 = 2(-\gamma_D k^2 + \gamma_B k)|B_{+}|^2 + 2\gamma_{\omega}^2 k^4 f(k,\eta) |v|^2
\end{equation}
\begin{equation}
\label{final2}
 \partial_\eta |B_{-}|^2 = 2(-\gamma_D k^2 - \gamma_B k)|B_{-}|^2 + 2\gamma_{\omega}^2 k^4 f(k,\eta) |v|^2
\end{equation}

Before we go for the numerical solution however, there are two things that need to be discussed. The evolution of the magnetic field depends on the comoving chemical potential 
$\Delta \mu$. The evolution of $\Delta \mu$ with $\eta$ has been discussed previously \cite{joyce,tashiro}. Instead of repeating the discussion, we just state that the comoving chemical potential 
evolves according to the following equation, 
\begin{equation}
\label{final3}
 \frac{d\Delta \mu}{d \eta} = - C_{\Delta} \alpha \int \frac{k dk}{2 \pi^2} [\partial_\eta |B_{+}|^2 - \partial_\eta |B_{-}|^2 ]
\end{equation}
Since we will be considering very high temperatures above the electroweak scale, the chirality flipping rate is ignored.   

The second thing is the possible existence of a thermal gradient in the plasma. As the left and the right handed particles are not in thermal equilibrium, their number densities 
will also be different. This generates a thermal current given by \cite{ahonen}, 
\begin{equation}
 J_T = - \alpha \sigma \nabla T + \frac{\sigma \nabla \mu}{e}. 
\end{equation}
However the thermal current does not make a difference to the evolution of the magnetic field. Since the Maxwell's equations lead to 
a curl of the current term and the curl of a gradient is zero hence the only effect of the hotter/colder particles will be to modify the chiral vorticity term. 
We now discuss the evolution of the magnetic field under the effect of the temperature dependent vorticity term.

\section{Numerical Results} 
\label{sec:results}
The equations Eq.(\ref{final1}), Eq.(\ref{final2}) and Eq.(\ref{final3}) have been evolved simultaneously and the magnetic field energy has been numerically 
obtained for the case of thermal equilibrium by Tashiro et al. \cite{tashiro}. 
They have found the magnetic field spectrum for longer lengthscales (lower values of $k$) to be proportional to $k^7$. 
In the Kolmogorov regime, they obtain a Gaussian peak. The Gaussian peak shifts to longer lengthscales and becomes sharper with evolution. We have used their 
results as a benchmark to test our numerical solutions.

In the presence of the hotter/colder particles, as we have discussed in the previous section \ref{sec:cve}, the primary change in the equations is in the factor  
$\gamma_\omega$. The second term of $\gamma_\omega$ depends on the temperature of the hotter/colder particles $T_{NT}$. 
\begin{equation}
 \gamma_\omega = \frac{e \Delta \mu^2}{4 \pi^2 \sigma} + \frac{eT_{NT}^2}{12 \sigma}
\end{equation}
 The factor of $\gamma_\omega$ now involves two terms. The first term
depends only on the chiral potential $\Delta\mu$, whereas the second term depends on the temperature of the hotter/colder particle. Since large 
temperature differences between the hotter/colder particle and the background may create hydrodynamical complications, we take the temperature of the 
hotter/colder particle to be $0.2 T, 0.5 T$ and $2 T$ where $T$ is the thermal background temperature. The  chiral potential $\Delta\mu$
is also taken to be of the order of the background temperature. We have then numerically solved the differential equations using the R software environment.

We consider the magnetic field to be generated at a temperature of $ 10^{10}$ GeV. The constants $\alpha = 1/137$, $e = \sqrt{4 \pi \alpha}$ and the comoving 
conductivity $\sigma = 70$. The chemical potential is of the order of the thermal background temperature $\Delta \mu \approx T$ and the temperature of the 
hotter/colder particles is taken to be $T_{NT} = 0.2, 0.5 , 2.0 $ in units of $T$ The thermal background temperature is $10^{9}$ GeV.  
\begin{figure}
\begin{center}
	\includegraphics[scale = 0.5, angle = 270]{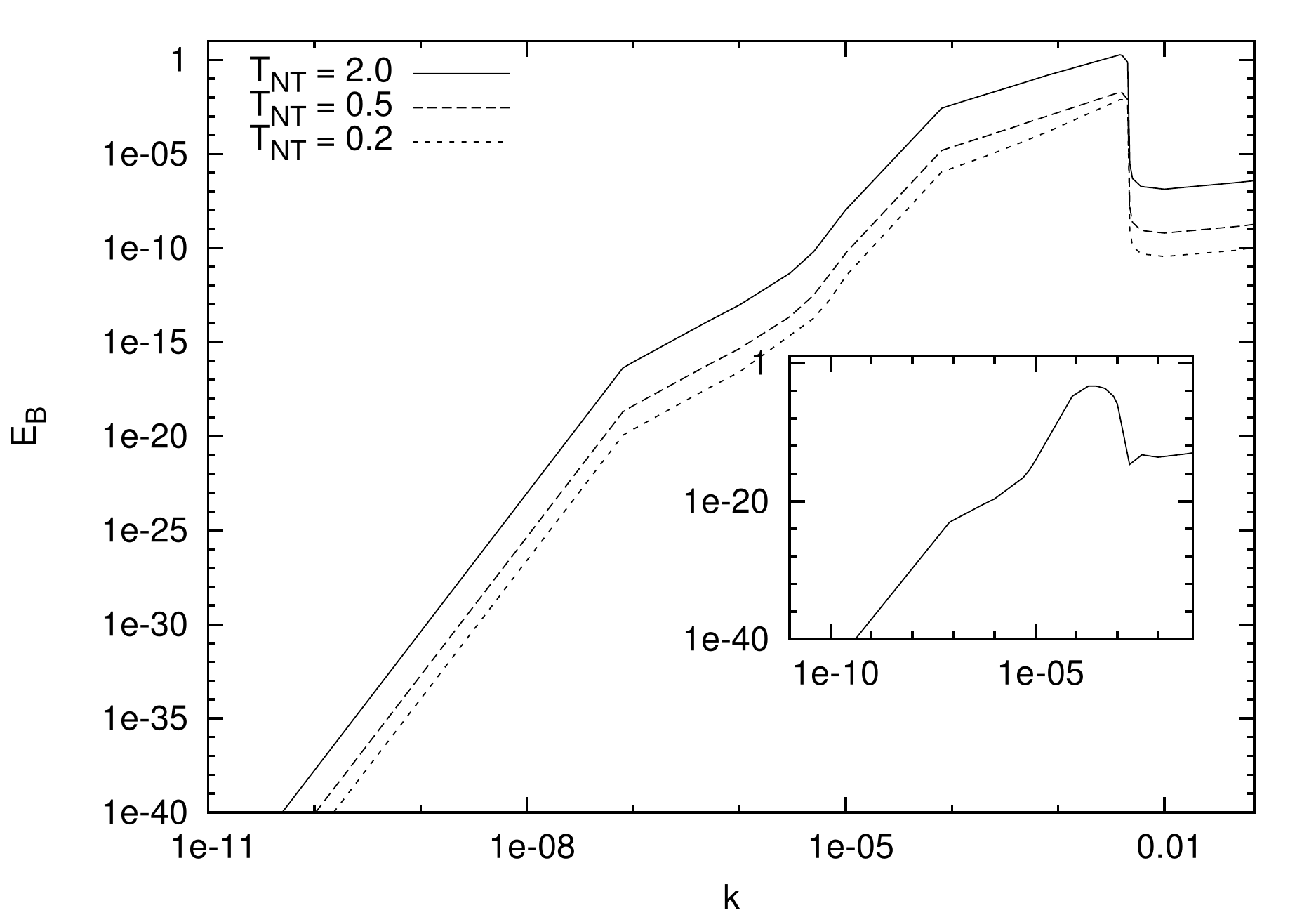}
	\caption{figure}{Plots of the magnetic energy spectrum in the presence of a hotter/colder particle when the background temperature is $T = 10^9 $ GeV. 
	The inset shows the evolution of the magnetic energy spectrum in the absence of the hotter/colder particles. The initial conditions for all the cases 
	remain the same.The initial value of $\Delta\mu$ is taken to be $1$.} 
\end{center}	
\end{figure}

Since we are interested in seeing the effect of the temperature dependent term in the vortical effect, we plot the spectrum of the magnetic field energy at a 
temperature $T = 10^{9}$ GeV. Evolving the spectrum to lower temperatures does not affect our results. We see that the evolution of the magnetic field at large 
lengthscales is unaffected by the presence of the hotter/colder particles. It is only in the Kolmogorov regime that 
we see a difference. The peak of the spectrum is skewed negatively. For small $k$ we obtain $E_B \propto k^7$ and a negatively skewed peak for $k > 10^{-4}$.

The presence of the particles at a different temperature
manifests in an increase in the vorticity factor. The form of the differential equation therefore changes. The magnetic field evolution equations have the general
form $\partial_\eta y + p(\eta) y = q(\eta)$ where $p(\eta)$ contains the diffusion and the chiral magnetic term, while $q(\eta)$ is the vorticity term. If $q(\eta)$ is set 
to zero, the solution of the equation will have a Gaussian form. In thermal equilibrium, for large wavenumbers (short lengthscales) the value of the vorticity 
term becomes negligible and it is possible to set  $q(\eta) \sim 0$. This leads to a Gaussian peak in the spectrum. But the presence of particles at a different temperature 
means that $q(\eta) \neq 0$ as it never becomes negligibly small.

An exact analytical solution to the differential equation is difficult to find but the numerical solution obtained can be fitted by an existing distribution. We
find that the family of beta distributions for various values of the parameters fit the solutions for different temperatures of the hotter/colder particles at the peak. 
The evolution of the magnetic field does not have the same structure for long and short wavelengths. The slope of the curve changes to give us the different regions. 
As mentioned before, for long wavelengths the impact of the hotter/colder particles are negligible. For small $k$ values we obtain the $k^7$ slope. Our interest therefore 
lies in the Kolmogorov regime. We fit the data in the peak region and find that it fits the beta distribution. In figure 2 we show only the 
fitted peak region.  
\begin{figure}
\begin{center}
	\includegraphics[scale = 0.5, angle = 270]{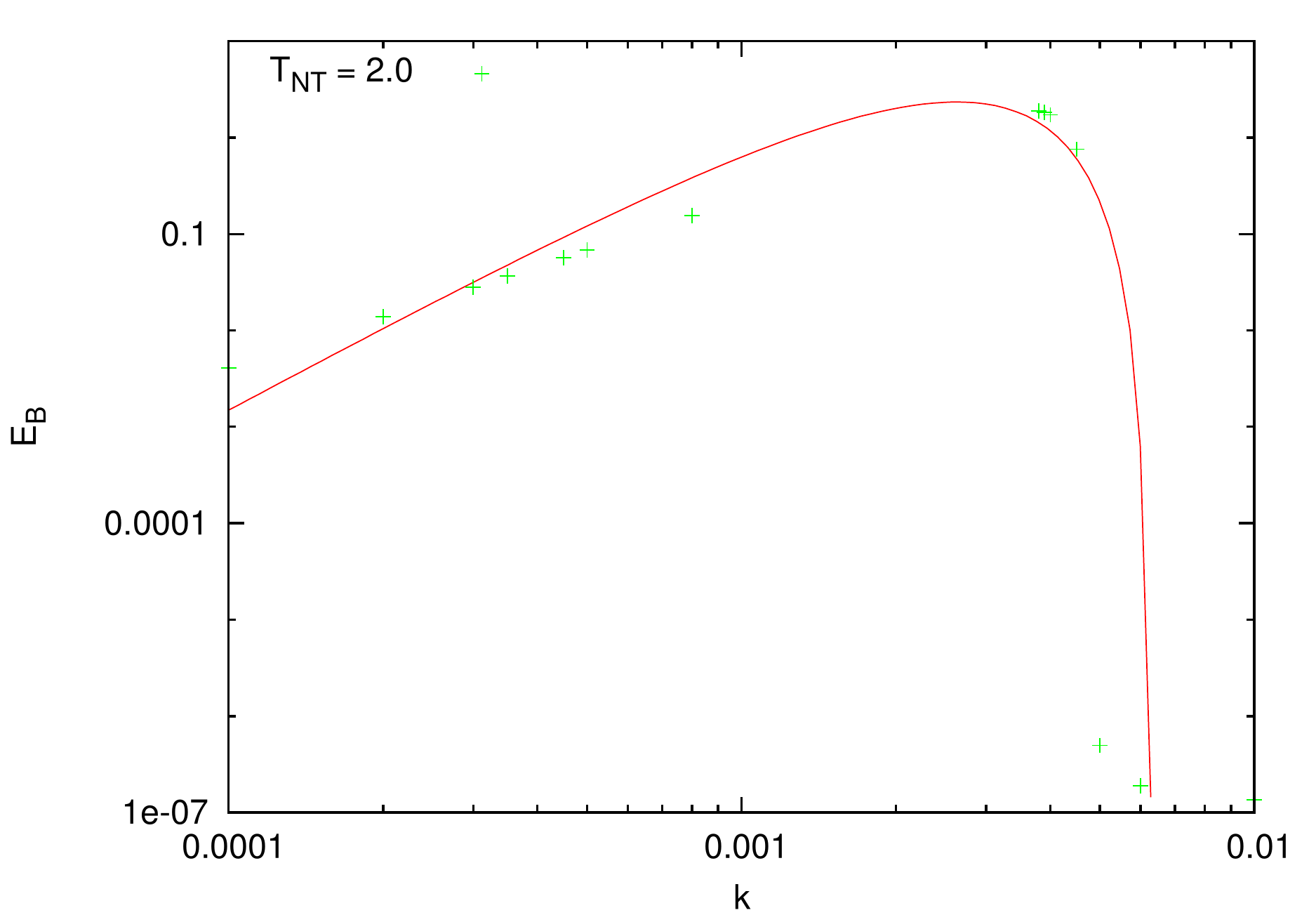}
	\caption{figure}{The peak region is fitted with a beta distribution. The figure shows the peak for $T_{NT} = 2.0$. It is possible to fit the other peaks too 
	for different values of the parameters for the beta distribution.}
\end{center}
\end{figure}

The beta distribution belongs to the family of exponential distributions and  has two free parameters. The different evolutionary graphs obtained for different temperatures of the thermal 
particles can be fitted be varying these two parameters. The generalized beta distribution has been used before to fit the non-Gaussianity of density fluctuations in 
a magnetized plasma in confined plasma devices \cite{labit}. Since experimentally, they were very good fits, there were attempts to provide a physical motivation 
of having a Beta distribution. The field became even more generalized when similar distributions were also found in completely unrelated fluid dynamical systems. 

Though we are discussing the evolution of a magnetized plasma, it is not confined by the magnetic field. Hence, the discussions and models for confined magnetic plasma 
devices would not be applicable in our case. The non-Gaussianity in the peak of the magnetic energy term can however be directly related to the strength of the 
vorticity term. The presence of the non-Gaussianity seems to indicate the presence of turbulence as we know that turbulence is a highly non-linear phenomenon. 
Turbulence in confined plasma's are essentially a combination of Gaussian fluctuations and non linearly interacting coherent structures 
\cite{guszejnov} which are driven by an external drive. Under a very weak drive, the system is not turbulent, only when the external drive 
is strong enough the effects of turbulence is seen in the system. In the case of the early universe, once the helical magnetic field develops, the chiral magnetic effect
starts dominating over the chiral vortical effect. The chiral vortical effect gradually becomes negligible. However, if for some external reason, the chiral vortical 
effect does not become negligible it may lead to turbulence in the magnetized plasma. It looks like the hotter/colder particles provide this extra drive and thus lead 
to the development of non-linearities in the magnetized plasma. 

One can also obtain an approximate analytical solution for the case when $\Delta\mu$ is kept constant. A general solution to the equation is then given by 
\begin{equation}
 |B^{\pm}|^2 \rightarrow e^{-2(\gamma_D k^2 \mp \gamma_B k)\eta}\int_{\eta_0}^\eta 2 
e^{2(\gamma_D k^2 \mp \gamma_B k)\eta'} \gamma_\omega^2 k^4 f(k,\eta')|v^2| d\eta' 
\end{equation}

The general solution is a combination of exponential functions, which is probably why we are able to fit the peak with a beta distribution. As mentioned before, 
the beta distribution belongs to the family of exponential distributions.
The magnetic fields obtained by solving these equations are very high magnetic fields. The magnitude of the fields increase with the increase in the temperature 
difference between the hotter/colder particles and the background. From the solution we notice that the $|B^{+}|^2 $ dominates over the  $|B^{-}|^2 $ modes. This is due to the presence 
of the exponential term outside the integral. Both the modes are enhanced by the increase in the $\gamma_{\omega}$ term. This term is within the integral. 
But, in the case of the $|B^{-}|^2 $ modes, there is a strong exponential decay due to the multiplicative term outside the integral and hence 
the effect of the hotter/colder particles is not visible here. For the  $|B^{+}|^2 $ modes, though there is an exponential decay due to the diffusion term, 
the chiral magnetic term does not contribute to the decay of the field. Hence the overall magnitude of the field is higher for the  $|B^{+}|^2 $  modes. 
Therefore, we plot only the $|B^{+}|^2 $  modes to show the enhancement of the magnetic field due to the hotter/colder particles. 

We have kept the 
temperatures of the hotter/colder particles close to the thermal background as several complications can arise if the temperature difference is high. In figure 3, 
we show the magnitude of the magnetic fields generated for different values of $k$. As expected, we find that at 
lower values of $k$, the magnetic field remains low, it becomes significant for $k > 10^{-4}$. This is the time when the chiral magnetic effect start to dominate.
However, as we have seen in the energy spectrum, the chiral magnetic effect cannot make the chiral vortical effect insignificant as long as the hotter/colder particles 
are there. 

In our analysis we do not look at the helicity spectrum as the vorticity term does not affect the helicity evolution.  
 The $\gamma_\omega$ term essentially gets cancelled out while calculating the helicity, so the evolution of the helicity depends only the chiral magnetic effect term 
 and the diffusion term. However as a check we plotted the helicity spectrum and found that it remained Gaussian even in the presence of the hotter/colder particles. 
\begin{figure}
\begin{center}
	\includegraphics{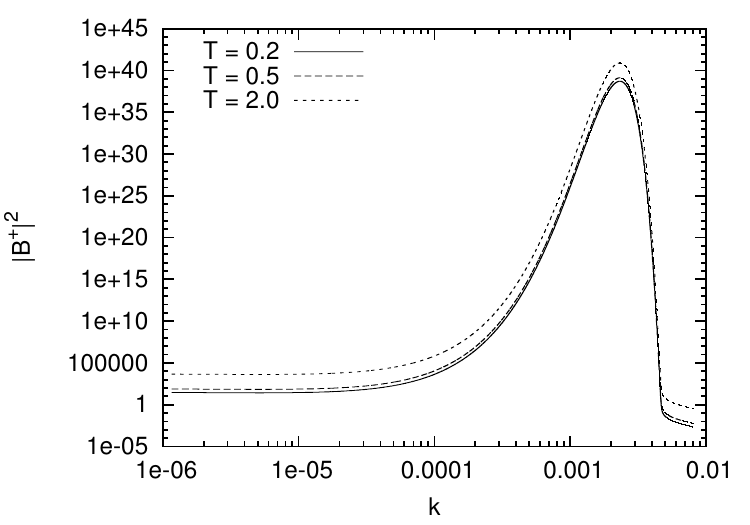}
	\caption{figure}{The approximate solution of the magnetic field evolution equation for a fixed value of chemical potential in the presence of a hotter/colder particles.} 
\end{center}
\end{figure}

Generally in most models of primordial magnetic field generation, the problem is that the magnitude of the generated field is not very large. As the universe expands, 
the magnetic energy therefore goes down. For the primordial field to be detectable today, a dynamo effect is required. This enhances the generated magnetic field.
We find that the magnitude of the generated helical magnetic field is enhanced if there are particles at temperatures different from the background temperature. 
We do not discuss the generation 
of the magnetic field. Tashiro et al \cite{tashiro} suggested that initial vorticity may generate the magnetic field through turbulence. We have shown that in the presence of 
hotter/colder particles, the vortical effect never becomes negligibly small compared to the chiral magnetic effect. Hence the generated magnetic fields would 
be larger in magnitude and therefore survive for a longer time. 

\section{Discussion}
\label{sec:discussion}
We have studied the evolution of a helical magnetic field in the presence of particles which have a temperature that is different from the background temperature. 
The particles are thus out of thermal equilibrium with the background 
plasma. We focus on the chiral magnetic effect and the chiral vorticity effect. While the former is unaffected by the presence of the hotter (or colder) particle, the latter 
is affected by it. Hotter/colder particles will also generate a temperature gradient but from Maxwell's equations we know that the current generated by a temperature 
gradient will not affect the evolution of the magnetic fields. So the main effect that we study is the chiral vortical effect. We find even when there is a strong 
magnetic field the chiral vortical effect does not become negligible compared to the chiral magnetic effect in the presence of hotter/colder particles. This leads to the 
turbulence in the plasma. 

It has been shown that for a helical magnetic field evolving in an expanding universe, the chiral magnetic effect leads to a Gaussian energy spectrum in the Kolmogorov 
regime. In the long lengthscales, where the white noise spectrum is generally assumed the energy spectrum has a $k^7$ dependence. We show that the temperature dependent 
term in the chiral vortical effect will lead to a energy spectrum which remains the same in the long lengthscales but changes to a non-Gaussian spectrum in the 
Kolmogorov regime. The peak of the spectrum can be fitted with a beta distribution. Since this kind of distributions have been used to fit density spectrums of 
turbulent confined magnetized plasma, we believe that the beta distribution indicates the presence of turbulence in the cosmological plasma. Turbulence can be 
generated in a plasma by the injection of energies through different sources. Here it is the hotter/colder particles, which enhances the vorticity term so that it 
can never become negligible compared to the other effects. 

Since the hotter/colder particles only affects the vorticity factor, the question arises whether a large 
chiral asymmetry would produce the same effects as the hotter particle ? A large chiral asymmetry first of all may not conform to the different observable 
bounds in the early universe \cite{mangano}. Moreover, the chemical potential actually decreases with time and reaches an aymptotic value. Even if the initial value of the 
chemical potential is large, it will go down to this value. This means that finally the chiral magnetic effect will dominate the evolution of the magnetic field 
again. The chemical potential is present in both the effects however the temperature of the particle only affects the chiral vorticity effect. Hence one can attribute 
the skewed non-Gaussian distribution to the presence of the temperature dependent term in the chiral vortical effect. 
There are various assumptions which have been taken in the course of this work. A better picture would arise from a complete solution of the magnetohydrodynamic 
equations including the advection term. We have shown that the chiral vorticity effect dominates over the chiral magnetic effect in the presence of a species of particles
which have a different temperature than the background. Such a case may occur for several nonthermal dark matter particles. We would like to point out that it is the temperature 
dependent term in the chiral vortical effect that is responsible for the results we have obtained. For a plasma, where the particles are in thermal equilibrium, this term is 
neglected. However it cannot be neglected when all the particles are not at the same temperature. Hence we discuss the plasma in the presence of a hotter/colder particle. 
Thus as long as the temperature term is present in the Chiral Vortical effect, we would get a non-Gaussian spectrum even in the absence of a hotter/colder particle. The 
non-Gaussianity is enhanced if there is a difference between the background and the particle temperature.

We do not discuss the generation of the primordial magnetic field in detail anywhere. We have assumed that it is generated by the vortical effect.
Even if it is generated by some other effect, our results would hold as long as there is some vorticity in the plasma.  
Our results do not depend on the initial value of the magnetic field. Once a helical magnetic field is generated as long as the vorticity term is 
present, our results indicate that the plasma dynamics will be turbulent. This also means that the magnitude of the magnetic fields are enhanced due 
to the presence of the hotter/colder particles. We find that the magnitude of the magnetic field depends on the temperature difference between the thermal background and the 
particle temperature. If this temperature 
difference is too large then other temperature effects will manifest themselves, hence we have considered only small but finite differences between the particles and the 
background.

{\bf Acknowledgements}

SS would like to acknowledge discussions with Abhishek Atreya, Jitesh Bhatt, N. Mahajan and H. Mishra. For the numerical evaluations we acknowledge suggestions and 
advice from Debasis Dan and the R - Project (free integrated suite of software for calculations, data manipulation and graphical display). 
TM is financially supported by the
SERB-DST Ramanujan Fellowship under Project no - SB/S2/RJN-29/2013.

\end{document}